\documentclass[aps,prd,preprint,superscriptaddress,tightenlines,nofootinbib]{revtex4}

\usepackage{graphicx}% Include figure files
\usepackage{dcolumn}% Align table columns on decimal point
\usepackage{bm}% bold math

\begin{document}

%\preprint line(s) will be ignored for PRL/PRD
%\preprint{CLEO Draft YY-NNA} % For paper draft CBX YY-NN -> Draft YY-NNA
%\preprint{CLEO CONF YY-NN}   % For conference papers
%\preprint{ICHEP ABSnnn}      % For conference papers
\preprint{CLNS 03/1823}       % for CLNS notes
\preprint{CLEO 03-07}         % for CLNS notes

\title{Measurement of the Charge Asymmetry in $B\to K^*(892)^\pm\pi^\mp$}
% for conference papers (ask CLEOAC for appropriate text)
%\thanks{Submitted to the 31$^{\rm st}$ International Conference on High Energy
%Physics, July 2002, Amsterdam}

%-------- INSERT HERE ------------
% Your author list goes here  REMOVE EVERYTHING to END INSERT and
% replace with your authorlist (ask cleoac).

\author{B.~I.~Eisenstein}
\author{G.~D.~Gollin}
\author{I.~Karliner}
\author{N.~Lowrey}
\author{C.~Plager}
\author{C.~Sedlack}
\author{M.~Selen}
\author{J.~J.~Thaler}
\author{J.~Williams}
\affiliation{University of Illinois, Urbana-Champaign, Illinois 61801}
\author{K.~W.~Edwards}
\affiliation{Carleton University, Ottawa, Ontario, Canada K1S 5B6 \\
and the Institute of Particle Physics, Canada}
% ...
%\author{(CLEO Collaboration)} %FOR PRD_SPECIAL_CHANGEME
\collaboration{CLEO Collaboration} %FOR PRL,CLNS
\noaffiliation

%-------- END INSERT ------------

%please hard code the date when you have a final draft and submit to CLEOAC
\date{April 22, 2003}

\begin{abstract} 
% Insert abstract here.
We report on a search for a $CP$-violating asymmetry in the charmless
hadronic decay $B\to K^*(892)^\pm\pi^\mp$, using 9.12 ${\rm fb}^{-1}$ of
integrated luminosity produced at $\sqrt{s}=10.58$ GeV and collected with
the CLEO detector.  We find
${\cal A}_{CP}(B\to K^*(892)^\pm\pi^\mp) = 0.26^{+0.33}_{-0.34}$(stat.)$^{+0.10}_{-0.08}$(syst.),
giving an allowed interval of $[-0.31,0.78]$ at the 90\% confidence level.
\end{abstract}
\pacs{13.20.He}
\maketitle

% Insert body of the text here.
The Standard Model predicts that $CP$-violating phenomena are governed solely
by the single imaginary parameter of the Cabibbo-Kobayashi-Maskawa
matrix~\cite{Kobayashi:fv} of complex quark couplings.  The first
observations of $CP$ violation in the neutral $B$ system were recently
reported~\cite{sin2beta}, and they have been interpreted widely as induced by
$B^0$-$\bar B^0$ mixing.  To date, direct $CP$ violation has only been
observed in the neutral kaon system~\cite{directKaonCPV}.
Direct $CP$ violation in a given decay requires contributions from two or more
amplitudes which differ in both $CP$-violating (weak) and $CP$-conserving
(strong) phases.
In the $B$ system, these conditions are expected to be met in some charmless
hadronic decays, and direct $CP$ violation can occur at sizeable levels,
depending on the magnitude of the strong phase
difference~\cite{acpTheory,kstpiAcpTheory}
or on the presence of new physics~\cite{newPhysics}.
Previous analyses, mainly focusing
on two-pseudoscalar final states, have not observed
direct $CP$ violation in these decays~\cite{cleoAcpPP,acpExperiment}.
In this Report, we present a search for direct $CP$ violation in the
vector-pseudoscalar decay $B\to K^*(892)^\pm\pi^\mp$.  We express the
difference between the decay rates for $\bar B^0\to K^*(892)^-\pi^+$ and
$B^0\to K^*(892)^+\pi^-$ in terms of an asymmetry, ${\cal A}_{CP}$, defined as
\begin{equation}
{\cal A}_{CP}\equiv
\frac{{\cal B}(\bar B^0\to K^*(892)^-\pi^+)-{\cal B}(B^0\to K^*(892)^+\pi^-)}
{{\cal B}(\bar B^0\to K^*(892)^-\pi^+)+{\cal B}(B^0\to K^*(892)^+\pi^-)}.
\end{equation}
We consider both $K^*(892)^\pm$ submodes, $K^*(892)^\pm\to K^0_S\pi^\pm$ and
$K^*(892)^\pm\to K^\pm\pi^0$ by analyzing the final states
$K^0_S\pi^\pm h^\mp$ and $K^\pm h^\mp\pi^0$, where $h^\mp$ denotes
a charged pion or kaon.  We perform a maximum likelihood fit in the
$K^0_S\pi^\pm h^\mp$ and $K^\pm h^\mp\pi^0$ Dalitz plots to distinguish
$B\to K^*(892)^\pm\pi^\mp$ from other intermediate resonances or
non-resonant three-body decays.
The $CP$-averaged branching fraction for $B\to K^*(892)^\pm\pi^\mp$
has been measured by the Belle~\cite{belle} and CLEO~\cite{cleo}
Collaborations, and the work described in this Report is an extension of that
previous CLEO analysis.

The data sample used in this analysis was produced in symmetric $e^+e^-$
collisions at the Cornell Electron Storage Ring (CESR)
and collected with the CLEO detector in two configurations, known as
CLEO II~\cite{cleonim} and CLEO II.V~\cite{iivnim}.
It comprises 9.1 ${\rm fb}^{-1}$ of integrated luminosity
collected on the $\Upsilon(4S)$ resonance, corresponding to $9.7\times 10^6$
$B\bar{B}$ pairs, of which $6.3\times 10^6$ were taken with CLEO II.V.
An additional 4.4 ${\rm fb}^{-1}$ collected below the $B\bar{B}$ production
threshold is used to study non-$B\bar{B}$ backgrounds.  Of this latter
luminosity, 2.8 ${\rm fb}^{-1}$ were collected with CLEO II.V.
The response of the experimental apparatus is studied with a detailed
GEANT-based~\cite{geant} simulation of the CLEO detector, where the
simulated events are processed in a fashion similar to data.

In CLEO II,
the momenta of charged particles are measured with a
tracking system consisting of a six-layer straw
tube chamber, a ten-layer precision
drift chamber, and a 51-layer main drift chamber, all operating
inside a 1.5 T superconducting solenoid.  The main drift chamber
also provides a measurement of specific ionization energy loss ($dE/dx$),
which is used for particle identification.
For CLEO II.V,
the six-layer straw tube chamber was replaced by a three-layer
double-sided silicon vertex detector, and the gas in the main 
drift chamber was changed from an argon-ethane to a helium-propane mixture.
Photons are detected with a 7800-crystal CsI electromagnetic calorimeter,
which is also inside the solenoid.
Proportional chambers placed at various depths within the steel return
yoke of the magnet identify muons.

Charged tracks are required to be well-measured and to satisfy 
criteria based on the track fit quality.  They must also be consistent with
coming from the interaction point in three dimensions. Pions and kaons are
identified by consistency with the expected $dE/dx$, and tracks
that are positively identified as electrons or muons are not allowed to form
the $B$ candidate.  We form $\pi^0$ candidates from
pairs of photons with invariant mass within 20 MeV/$c^2$ (approximately
2.5 standard deviations ($\sigma$)) of the known $\pi^0$ mass.
These candidates are then kinematically fitted with their masses
constrained to the known $\pi^0$ mass.  We also require the $\pi^0$
momentum to be greater than 1 GeV/$c$
to reduce combinatoric background from low-momentum $\pi^0$ candidates.
$K^0_S$ candidates are selected from pairs of tracks with invariant
mass within 10 MeV/$c^2$ or $2.5\sigma$
of the known $K^0_S$ mass.  In addition, $K^0_S$ candidates are required
to originate from the beam spot and to have well-measured displaced
decay vertices.

We identify $B$ meson candidates by their invariant mass and the total
energy of their decay products.  We calculate a beam-constrained mass by
substituting the beam energy ($E_b$) for the measured $B$
candidate energy: $M \equiv\sqrt{E_b^2 - {\mathbf p}_B^2}$, where
${\mathbf p}_B$ is the $B$ candidate momentum.  Performing this substitution
improves the resolution of $M$ by one order of magnitude, to about
3 MeV/$c^2$.  We define $\Delta E\equiv E_1 + E_2 + E_3 - E_b$, where
$E_1$, $E_2$, and $E_3$ are the energies of the $B$ candidate daughters.
For final states with a $K^0_S$ and two charged tracks, the $\Delta E$
resolution is about 20 MeV for CLEO II and 15 MeV for CLEO II.V.  A $\pi^0$
in the final state degrades this resolution by approximately a factor of two.
$\Delta E$ is always calculated assuming the $h^\mp$ is a pion.
Therefore, the $\Delta E$ distribution for pions is centered at zero,
while that for kaons is shifted by at least $-40$ MeV.
We accept $B$ candidates with $M$ between 5.2 and 5.3 GeV/$c^2$ and with
$|\Delta E|$ less than 300 MeV for $K^\pm h^\mp\pi^0$ and 200 MeV for
$K^0_S\pi^\pm h^\mp$.  This
region includes the signal region and a high-statistics sideband for background
normalization.  We reject candidates that are consistent with the
exclusive $b\to c$ transitions $B\to D\pi$, where $D\to K\pi$,
and $B\to\psi K^0$, where $\psi\to\mu^+\mu^-$ and the muons are misidentified
as pions.

The main background in this analysis arises from $e^+e^-\to q\bar q$,
where $q = u,d,s,c$.  To suppress this background, we calculate the
angle $\theta_{\rm sph}$ between the sphericity axis~\cite{sphericity}
of the tracks and showers
forming the $B$ candidate and that of the remainder of the event.  Because of
their two-jet structure, continuum $q\bar q$ events peak strongly at
$|\cos\theta_{\rm sph}|=1$, while the more isotropic $B\bar{B}$
events are nearly
flat in this variable.  By requiring $|\cos\theta_{\rm sph}|<0.8$, we reject
83\% of the continuum background while retaining 83\% of
signal $B$ decays.  Additional separation of signal from $q\bar q$ background
is provided by a Fisher discriminant~\cite{Fisher} ${\cal F}$
formed from eleven
variables: the angle between the sphericity axis of the candidate and the
beam axis, the ratio of Fox-Wolfram moments $H_2/H_0$~\cite{r2}, and the
scalar sum of the visible momentum in nine $10^\circ$ angular bins around
the candidate sphericity axis.  In the likelihood fit, we also make use of the
angle between the $B$ candidate momentum and the beam axis, $\theta_B$.
Angular
momentum conservation causes $B$ mesons produced through the $\Upsilon(4S)$ to
exhibit a $\sin^2\theta_B$ dependence, while candidates from continuum
are flat in $\cos\theta_B$.

In both the $K^0_S\pi^\pm h^\mp$ and $K^\pm h^\mp\pi^0$ topologies, the
$h^\mp$ refers to the faster of the two tracks, which typically has momentum
above 1 GeV/$c$.  Because $dE/dx$ still provides limited separation of pions
and kaons above 1 GeV/$c$, we make use of the $dE/dx$ information in the
likelihood fit.  In $B\to K^*(892)^\pm\pi^\mp$ decays, this higher-momentum
track is the one that recoils from the $K^*(892)^\pm$, more than 99.99\% of the
time.  Its charge uniquely distinguishes $\bar B^0\to K^*(892)^-\pi^+$ from
$B^0\to K^*(892)^+\pi^-$.  Thus, the charge asymmetry ${\cal A}_{+-}$,
formed using the charge of this higher-momentum track, is essentially the same
as ${\cal A}_{CP}$.

Our loose selection criteria result in samples consisting primarily of
background events and containing 11893 candidates for $K^0_S\pi^\pm h^\mp$ and
28589 for $K^\pm h^\mp\pi^0$.  To extract yields and $CP$ asymmetries, we
perform an unbinned maximum likelihood fit  using the observables
$M$, $\Delta E$, ${\cal F}$, $\cos\theta_B$, the two Dalitz plot variables
in each topology, and the $dE/dx$ of the $h^\mp$ (the faster of the two primary
tracks).  At high momentum, charged pions and kaons are statistically separated
by their $dE/dx$ and by $\Delta E$, each of which provides discrimination at
the $2.0\sigma$ level ($1.7\sigma$ for CLEO II), and we fit for both $\pi$ and
$K$ hypotheses simultaneously.  Charged pions and kaons with momentum below
1 GeV/$c$ are cleanly identified by $dE/dx$ consistency at the $3\sigma$ level.
The free parameters in the fit are yields ($N$) summed over charge states,
$N_{h^+}+N_{h^-}$, and charge asymmetries,
${\cal A}_{+-}\equiv(N_{h^+}-N_{h^-})/(N_{h^+}+N_{h^-})$.

The probability for a candidate to be consistent with a given component is
the product of the probability density function (PDF) values for each of
the input variables (neglecting
correlations).  The likelihood for each candidate is the sum of probabilities
over
the components in the fit, with relative weights determined by maximizing the
total likelihood of the sample, which is given by the following expression:
\begin{equation}
{\cal L} = \prod_{i=1}^{\rm candidates} \left\{
\sum_{j=1}^{\rm components} \left[ f_j (1\pm{\cal A}_{+-}^j)
\prod_{k=1}^{\rm variables} {\cal P}_{ijk} \right] \right\},
\end{equation}
where the $\pm$ refers to the charge of $h^\pm$ in each candidate.
The ${\cal P}_{ijk}$ are the per candidate PDF values, and the $f_j$ and
${\cal A}_{+-}^j$ are the free parameters optimized by the fit. 
The products $f_j(1\pm{\cal A}_{+-}^j)/2$ are constrained to sum to the
fraction of candidates in the fit with the appropriate charge of $h^\pm$.
Since the PDFs are normalized to unit integral over the fit domain,
the $f_j$ can be interpreted as component fractions.
The parameters of the $dE/dx$ PDFs are measured from $D\to K^\pm\pi^\mp$ decays
in data. For all other variables, the signal and the background $b\to c$ PDFs
are determined
from high-statistics Monte Carlo samples, and the continuum PDFs are determined
from data collected below the $B\bar{B}$ production threshold.  
The impact of correlations among the input variables is reduced by determining
the PDFs as a function of the event location in the Dalitz plot,
for coarse bins in the $M^2(K\pi)$-$M^2(\pi\pi)$ plane.
We use Monte Carlo simulation to estimate the
systematic uncertainty associated with neglecting any remaining correlations.

Events from $B\to K^0_S\pi^\pm h^\mp$ and $B\to K^\pm h^\mp\pi^0$, including
$B\to K^*(892)^\pm\pi^\mp$, are modeled in the fit as follows.
We consider various $B$ decay channels with
intermediate resonances ($K^*(892)$, $K^*_0(1430)$, $\rho(770)$, and
$f_0(980)$) as well as non-resonant phase space decay.  The Dalitz plot
PDFs include our knowledge of the
helicity structure in these decays.  We neglect interference among these
signal processes and assign a systematic uncertainty estimated from Monte
Carlo simulation.
The decays $B\to {\cal K}^\pm h^\mp$, where ${\cal K}^\pm$ denotes $K^*(892)^\pm$ or
$K^*_0(1430)^\pm$, are accessible through different ${\cal K}^\pm$ submodes in both
the $K^0_S\pi^\pm h^\mp$ and $K^\pm h^\mp\pi^0$ topologies.  
To maximize our sensitivity to these decays, we
fit these two topologies simultaneously, with the branching fraction and
charge asymmetry
for each ${\cal K}^\pm h^\mp$ decay constrained to be equal in its two ${\cal K}^\pm$
submodes, which are related by isospin.

The charge symmetry of the CLEO detector, the track reconstruction software,
and the $dE/dx$ measurement has previously been
verified~\cite{cleoAcpPP,Coan:1999kh}.
The charge asymmetries of the fit samples are $0.010\pm 0.009$ for
$B\to K^0_S\pi^\pm h^\mp$ and $0.006\pm 0.006$ for $B\to K^\pm h^\mp\pi^0$.
Detection efficiencies and crossfeed among the charge-summed signal yields are
measured from Monte Carlo simulated events and are accounted for in the fit
for ${\cal A}_{+-}$. We find the charge asymmetry of the detection efficiencies
to be consistent with the expected null result.
Crossfeed among different charge states is not included
in the fit, and its effect is estimated with Monte Carlo simulation.

We perform the fit with differing combinations of intermediate resonant and
non-resonant states, with up to twelve signal components.  The fitted value
of ${\cal A}_{CP}$ does not depend heavily on the number of signal components
in the fit, and we include a systematic uncertainty for these variations.
We also allow for four background components: pion and kaon hypotheses for
$h^\pm$ for continuum background and background from $b\to c$ decays.  We do
not fit for charge asymmetries in the background components, but we measure
them to be consistent with zero.  The $B\to K^*(892)^\pm\pi^\mp$ event yields
were measured to be~\cite{cleo} $12.6^{+4.6}_{-3.9}$ for
$K^*(892)^\pm\to K^0_S\pi^\pm$ and $6.1^{+2.2}_{-1.9}$ for
$K^*(892)^\pm\to K^\pm\pi^0$ with a combined statistical significance of
$4.6\sigma$.
In the fit, these yields are corrected for efficiency and crossfeed from other
modes, and the $CP$ asymmetry in $B\to K^*(892)^\pm\pi^\mp$ is measured to be
${\cal A}_{CP} = 0.26^{+0.33}_{-0.34}$$^{+0.10}_{-0.08}$, where the
uncertainties are statistical and systematic, respectively.  The dominant
contributions to the latter are statistical uncertainties in the PDFs and
variations in the fitting method.

\begin{figure}
%this works for computers connected to the LEPP network.
%copy the figure file from Cornell if you want to try this elesewhere.
\includegraphics*[width=3.75in]{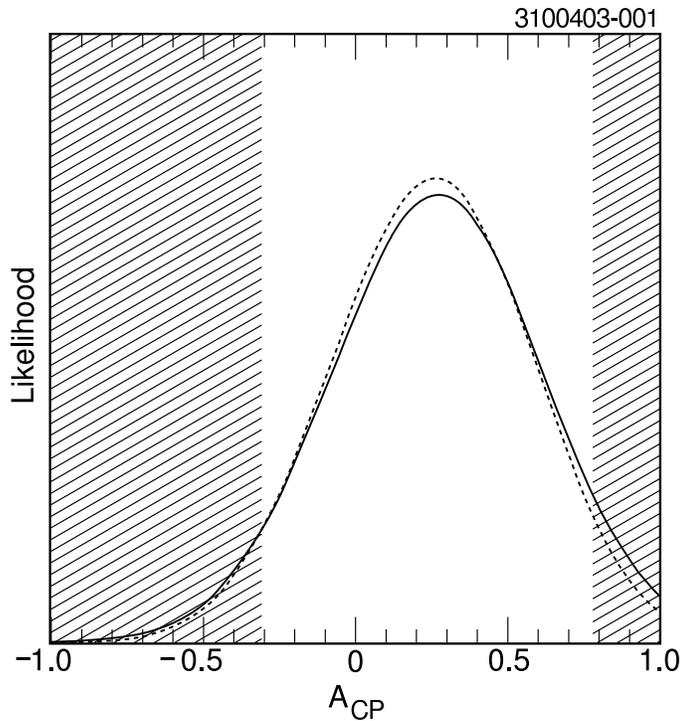}
\caption{Likelihood as a function of ${\cal A}_{CP}$
before (dashed) and after (solid) including systematic uncertainties.  The
hatched regions each contain 5\% of the integrated area in the physical
region and are excluded at the 90\% confidence level.}
\label{fig:lVsAcp}
\end{figure}

We determine the dependence of the likelihood function on ${\cal A}_{CP}$ by
repeating the fit at several
fixed values of ${\cal A}_{CP}$. By convoluting this function with the
systematic uncertainties and integrating the resultant curve in the physical
region, we construct
a 90\% confidence level interval of $-0.31 < {\cal A}_{CP} < 0.78$, where
the excluded regions on both sides each contain 5\% of the integrated area.
Figure~\ref{fig:lVsAcp} shows the likelihood function given by the fit and the
effect of including systematic uncertainties.

In summary, we have measured the $CP$ asymmetry in $B\to K^*(892)^\pm\pi^\mp$
using a simultaneous maximum likelihood fit to the $B\to K^0_S\pi^\pm h^\mp$
and $B\to K^\pm h^\mp\pi^0$ topologies.  We obtain the value
${\cal A}_{CP} = 0.26^{+0.33}_{-0.34}$$^{+0.10}_{-0.08}$,
which is consistent with the theoretical predictions~\cite{kstpiAcpTheory}
of $-0.19$ to 0.47. 
We also establish a 90\%
confidence level interval of $[-0.31,0.78]$.

%Example of how to insert eps files as figures

% CURRENT acknowledgements go here...
% download from the CLEO website 
% http://www.lns.cornell.edu/restricted/CLEO/analysis/ac_help/ack.html
We gratefully acknowledge the effort of the CESR staff 
in providing us with
excellent luminosity and running conditions.
M. Selen thanks the Research Corporation, 
and A.H. Mahmood thanks the Texas Advanced Research Program.
This work was supported by the 
National Science Foundation, 
and the
U.S. Department of Energy.

\end{document}